\documentclass[prd,a4paper,superscriptaddress,nofootinbib,10pt]{revtex4} 
\usepackage{graphicx}
\usepackage{amssymb}
\usepackage{amsmath}
\usepackage{lscape}
\usepackage{mathrsfs}
\usepackage{textcomp}
\usepackage{epsfig}
\usepackage{slashed}
\usepackage{comment}

\begin{document}
\title{Vacuum energy and the cosmological constant problem in $\kappa$-Poincar\'e invariant field theories}

\author{Tajron Juri\'c}
\email{tjuric@irb.hr}
\affiliation{Rudjer Bo\v{s}kovi\'c Institute, Bijeni\v cka  c.54, HR-10002 Zagreb, Croatia}

\author{Timoth\'e Poulain}
\email{timothe.poulain@th.u-psud.fr}
\affiliation{Laboratoire de Physique Th\'eorique, B\^at.\ 210
CNRS and Universit\'e Paris-Sud 11,  91405 Orsay Cedex, France}

\author{Jean-Christophe Wallet}
\email{jean-christophe.wallet@th.u-psud.fr}
\affiliation{Laboratoire de Physique Th\'eorique, B\^at.\ 210
CNRS and Universit\'e Paris-Sud 11,  91405 Orsay Cedex, France}

\date{\today}

\begin{abstract}

We investigate the vacuum energy in $\kappa$-Poincar\'e invariant field theories. It is shown that for the equivariant Dirac operator one obtains an improvement in UV behavior of the vacuum energy and therefore the cosmological constant problem has to be revised. 

\end{abstract}

\maketitle

\section{Introduction}

The history of the cosmological constant starts with Albert Einstein, where in his first paper on the application of  general relativity (GR) to cosmology \cite{einstein}, he aimed to construct a static universe with a finite average density of matter. In doing so he concluded that one needs to introduce an additional term to the gravitational equation of motion, namely the cosmological constant $\lambda$.\\

 One can visualize $\lambda$ as the curvature of the empty space. However, in GR curvature is connected with energy, momentum and pressure of matter. If we transfer the $\lambda$-term to the right hand side of the Einstein equation
\begin{equation}
R_{\mu\nu}-\frac{1}{2}g_{\mu\nu}R=8\pi G T_{\mu\nu}-g_{\mu\nu}\lambda
\end{equation}
one sees that the empty space produces the same gravitational field as when the space contains matter (or more precisely an ideal fluid) with density $\rho_\lambda=\frac{\lambda}{8\pi G}$ and pressure $p_\lambda=-\rho_\lambda$. In this sense we can speak of an energy density and pressure of the vacuum. The quantities $p_\lambda$ and $\rho_\lambda$ are the same in all coordinate systems (Lorentz-transformed moving relative to one another), so they will never appear in experiments with elementary particles, neither in atomic or molecular physics, since the cosmological constant term will always cancel out in the energy conservation law \cite{zeldovich}. Cosmological constant term only reveals himself in the gravitational phenomena on a large scale, i.e. cosmology.\\

The first problem with vacuum energy producing a gravitational effect was identified by W. Nernst \cite{nernst}, and later it developed into what is nowadays known as the \textit{cosmological constant problem}. More precisely, the cosmological constant problem is the disagreement between the observed value of the cosmological constant (vacuum energy density) and the theoretically large value of zero-point energy obtained within the quantum field theory (QFT). Namely, depending on the Planck energy cutoff and other factors, the discrepancy can go up to 120 orders of magnitude, being  ``the worst prediction in the history of physics''. For more details on the history and development on the attempts of solving the cosmological constant problem, the reader is advised to look at some of the following papers \cite{Padmanabhan:2002ji, Carroll:2000fy, Weinberg:1988cp, Sola:2013gha}.\\

The nature of the vacuum energy in QFT highly depends on the UV structure of the theory under question \cite{Nesterov}. It is determined that the classical (smooth) notion of spacetime is no longer adequate at Planck scale in order to reconcile GR with quantum mechanical axioms \cite{dop1, dop2, ahluwalia}. Properties of spacetime at the Planck scale are very different from what we expect from GR. Namely, various models including string theory \cite{string}, loop quantum gravity \cite{loop}, and  noncommutative (NC) geometry \cite{connes}, suggest that the spacetime might have a certain discrete structure at the quantum gravity scale. Combining together GR and quantum uncertainty principle one can predict a very general class of NC spaces \cite{dop1, dop2, ahluwalia}. Once the NC structure of spacetime is assumed, NC field theories  arises very naturally \cite{Douglas:2001ba, Szabo:2001kg, Wallet:2007em}. These NC field theories can have features like non-locality, UV/IR mixing and completely different UV behavior from their commutative counterparts \cite{Grosse:2012uv, Grosse:2004yu, Grosse:2003nw, Wallet:2016ilh, Gere:2015ota, Juric:2017bpr, Juric:2016cfp, Gere:2013uaa, Vitale:2012dz}.\\
 
Motivated by recent papers \cite{andjelo, andmim} we want to further investigate the nature of vacuum energy in NC field theories. In \cite{andjelo} the hypothesis that in models with an explicit breaking of Lorentz symmetry one can cure or improve the cosmological constant problem was tested. The result found there is that for certain Lorentz violating models one indeed obtains an improvement, while for the NC field theories examined there the result was negative. Furthermore, in the followup paper \cite{andmim} it was found that by examining appropriate measure on Hilbert space one can indeed obtain improvement of UV behavior of the vacuum energy in NC field theories. In the present work we will be interested in a very specific, but widely known Lie algebraic type of NC space, the so called $\kappa$-Minkowski space \cite{kappa1, kappa2, majid} and the field theories on it.\\

NC field theories on $\kappa$-Minkowski space were widely investigated \cite{Agostini:2004cu, Agostini:2003vg, Agostini:2006nc, Dimitrijevic:2003wv, Dimitrijevic:2003pn, Dimitrijevic:2005xw, Dimitrijevic:2014dxa, Meljanac:2007xb, Meljanac:2010ps, Harikumar:2011um, Meljanac:2011cs, Grosse:2005iz, Juric:2015jxa, Juric:2015hda}, but only recently the full investigation of the quantum properties of the 2-point and 4-point functions was explored in \cite{walletkappa, walpol}. In \cite{walletkappa}  a natural $\star$-product for $\kappa$-Minkowski space was used to investigate various classes of $\kappa$-Poincar\'e invariant scalar field theories with quartic interactions. $\kappa$-Poincar\'e symmetry induces a twisted trace which defines a KMS weight for the NC $\mathcal{C}^{*}$-algebra that models the $\kappa$-Minkowski space. It was found that in all examined  NC field theories, the twist generates different one-loop contributions to the 2-point functions which are at most UV linearly divergent. Some of the theories are free of UV/IR mixing. The  one-loop contribution to the 4-point function   is even found UV finite \cite{walpol} for some models whose kinetic operators are related to the square of the Dirac operator involved in the construction of an equivariant spectral triple \cite{dandrea} aiming to encode the geometry of $\kappa$-Minkowski space. This is partly due to the large spatial momentum damping in the propagator which decays as $1/p^{4}$. This strong decay of the propagator at large (spatial) momenta could ensure the perturbative renormalizability to all orders.\\

In this paper we will analyze the vacuum energy in $\kappa$-Poincar\'e invariant field theories for three different choices of kinetic operators: the Casimir operator \cite{majid}, the square of modular Dirac operator \cite{modular}, and square of equivariant Dirac operator \cite{dandrea}. Exploiting the $\kappa$-Poincar\'e symmetry and properties of the $\star$-product for $\kappa$-Minkowski space we formulate action functional and show that in the case of the square of equivariant Dirac operator one obtains a significant improvement in the UV behavior of the vacuum energy. Therefore one needs to revise the cosmological constant problem accordingly.\\

The paper is organized as follows. In section II we outline the standard computation of vacuum energy using partition function and heat kernel. We present the construction of the $\kappa$-Poincar\'e invariant field theories in section III. In section IV the vacuum energy for three different choices of kinetic operators is calculated.  We conclude with some final remarks in section V. Technical details on construction of the $\star$-product and appropriate Hilbert product are given in the Appendices.

\section{Vacuum energy and cosmological constant}

Consider a Euclidean bosonic field theory defined by the partition function
\begin{equation}
Z=\int \mathcal{D}[\phi]\ e^{-\mathcal{S}[\phi]}
\end{equation}
where
\begin{equation}\label{SF}
\mathcal{S}[\phi]=\int d^4 x (\phi F\phi)(x)=\int d^4 x d^4 y\ \phi(x)\phi(y)F(x,y)
\end{equation}
is the action functional and $F(x,y)=\int d^4 p \ e^{ip(x-y)}\tilde{F}(p)$ is the kernel of a generalized kinetic operator $F$ (and its Fourier transform). Then the partition function $Z$ is simply given by  $Z=(\text{Det} F)^{-1/2}$ and the effective action $W=\ln Z$ can be written in terms of the heat kernel \cite{Nesterov, vasi}
\begin{equation}\label{W}
W=-\frac{1}{2}\int^{\infty}_{\frac{1}{M^2}}\frac{ds}{s}H(s)
\end{equation}
where $s$ is a real parameter, $H(s)=\text{Tr}\ e^{-sF}$ is the trace of the heat kernel. The cutoff $1/M^2$ is introduced because in the standard field theory the effective action $W$ is usually divergent. Recall that the heat kernel $H(s; x, x^\prime)=(x|e^{-sF}|x^\prime)$ is defined as a solution of the heat equation
\begin{equation}
\left(\frac{\partial}{\partial s} +F\right)H(s; x, x^\prime)=0, \quad H(0; x, x^\prime)=\delta(x-x^\prime).
\end{equation}
It is usually more convenient to perform calculations in the momentum space, so that the trace of the heat kernel becomes
\begin{equation}\label{H}
H(s)=\frac{V}{(2\pi)^4}\int^{\infty}_{-\infty}d^4 p\ e^{-s\tilde{F}(p)}
\end{equation}
where $V$ is the volume of the spacetime. The vacuum energy density $\rho_{vac}$ is defined as 
\begin{equation}
\rho_{vac}=-\frac{W}{V}
\end{equation}
and it is often identified with the cosmological constant $\lambda$, but the relation between them is given by $\lambda=8\pi G\rho_{vac}$.\\

For the standard massless bosonic quantum field theory in 4-d space we have $F=\partial_{\mu}\partial^{\mu}$, so that the heat kernel becomes 
\begin{equation}
H(s)=\frac{V}{(4\pi s)^2}
\end{equation}
and the vacuum energy is given by 
\begin{equation}
\rho_{vac}(M)=\frac{M^4}{64\pi^2}
\end{equation}
If we consider the cutoff $M$ to be proportional to the Planck mass $M_{Pl}\approx 10^{18}$GeV, then we have a huge discrepancy of almost 120 orders of magnitude between the vacuum energy of the quantum field theory and the observed cosmological constant $\rho^{ob}_{vac}\approx 10^{-47}$GeV$^{4}$, leading to the ``worst prediction of theoretical physics''. However, this is just an estimate and no physical quantity can be cut-off dependent, since this dependance should be absorbed in the renormalized cosmological costant \cite{Grensing}.
Since there are indications that a change in the UV behavior of the kinetic operator  could drastically affect the structure of the vacuum energy \cite{Nesterov, andjelo, andmim}, we will be interested in investigating $\kappa$-Poincar\'e invariant field theories because they are characterized by kinetic operators which exhibit significant modifications in the UV regime.  \\

\section{$\kappa$-Poincar\'e invariant field theories}

\subsection{$\star$-product for $\kappa$-Minkowski space}
$\kappa$-Minkowski space \cite{kappa1}-\cite{majid} can be viewed as the universal enveloping algebra of the Lie algebra $\mathfrak{g}$ defined by
\begin{equation}
[x_0, x_i]=\frac{i}{\kappa}x_i, \quad [x_i , x_j]=0, \quad i=1,2,3
\end{equation}
where $\kappa>0$ is a deformation parameter of mass dimension.   
Furthermore, one can define a Lie group $\mathcal{G}$ by exponentiation\footnote{Here note that we chose the so called  time-to-the-right ordering.  Other orderings could be chosen (weyl-symmetric or time to the left) by simply exponentiating the Lie algebra via different ordering but this corresponds to different choices of coordinates system on the group manifold. Accordingly, the expression of the star product will change but, keeping in mind that we work within the framework of harmonic analysis, performing a change of coordinates at the level of (11) will also imply a modification of the (haar) measure appearing in the  eq.(13) for the star product. }
\begin{equation}
W(p_0,p_i)=e^{ip_{i}x_{i}}e^{ip_0 x_0}
\end{equation}
from which follows the group law
\begin{equation}
W(p_0,p_i)W(q_0,q_i)=W(p_0+q_0, p_i+e^{-\frac{p_0}{\kappa}}q_i)
\end{equation}
A convenient presentation of  $\kappa$-Minkowski space is provided by a mere adaptation of the Wigner-Weyl quantization scheme giving rise to the celebrated Moyal product by replacing the Heisenberg group with $\mathcal{G}$. This latter approach, whose main steps are recalled in the Appendix (see also \cite{sitarz, walletkappa}), leads to the following expressions for the $\star$-product and the adjoint
\begin{equation}\begin{split}
(f\star g)(x)&=\int dp_0 dy_0\  e^{-iy_0 p_0}f(x_0+y_0, x_i)g(x_0, e^{-\frac{p_0}{\kappa}}x_i)\\
f^{\dagger}(x)&=\int dp_0 dy_0\  e^{-iy_0 p_0}\bar{f}(x_0+y_0,e^{-\frac{p_0}{\kappa}}x_i)
\end{split}
\end{equation}
for any functions $f,g\in \mathcal{F}(S_{c})$. Here $S_c$ is the set of Schwartz functions on $\mathbb{R}^4$ with compact support in the first variable and $\mathcal{F}$ is the Fourier transform. For all physical purposes it will be sufficient to identify the algebra modeling $\kappa$-Minkowski $\mathcal{M}_\kappa$ with the algebra $\mathcal{F}(S_c)$. \\

\subsection{$\kappa$-Poincar\'e algebra and  invariant action}

In the present study, we require the action functional $\mathcal{S}_\kappa$ to be $\kappa$-Poincar\'e invariant which is a reasonable assumption regarding the important role played by the (classical) Poincar\'e symmetries in standard QFT together with the fact that $\kappa$-Poincar\'e algebra can be viewed as describing the quantum symmetries of the $\kappa$-Minkowski space-time \cite{kappa1}-\cite{majid}. These symmetries are encoded algebraically by the Hopf algebra of $\kappa$-Poincar\'e algebra $\mathcal{P}_{\kappa}$. One way to present $\mathcal{P}_{\kappa}$ is by using the 11 elements: $P_i$ are the momenta, $N_i$ are boosts, $M_i$ are rotations, and $\mathcal{E}$ is the shift operator related to time translations $\mathcal{E}=e^{-\frac{P_0}{\kappa}}$. They all satisfy the following commutation relations
\begin{equation}\begin{split}
[M_i,M_j]=i\epsilon_{ijk}M_k, &\quad [M_i, N_j]=i\epsilon_{ijk}N_k, \quad [N_i, N_j]=-\epsilon_{ijk}M_k\\
[M_i, P_j]=i\epsilon_{ijk}P_k, &\quad [P_i, \mathcal{E}]=[M_i, \mathcal{E}]=0, \quad [N_i, \mathcal{E}]=\frac{i}{\kappa}P_i\mathcal{E}\\
[N_i, P_j]&=-\frac{i}{2}\delta_{ij}\left(\kappa(1-\mathcal{E}^2)+\frac{1}{\kappa}p^{2}_{k}\right)+\frac{i}{\kappa}p_i p_j
\end{split}\end{equation}
$\kappa$-Poincar\'e algebra can be endowed by a  Hopf algebra structure. Let us define the coproduct $\Delta: \mathcal{P}_\kappa \rightarrow \mathcal{P}_\kappa\otimes\mathcal{P}_\kappa$, counit $\varepsilon:\mathcal{P}_\kappa\rightarrow\mathbb{C}$ and antipode $S:\mathcal{P}_\kappa\rightarrow\mathcal{P}_\kappa$ satisfying 
\begin{equation}\begin{split}\label{cop}
\Delta P_0=P_0\otimes 1 +1\otimes P_0, &\quad \Delta P_i=P_i\otimes 1+\mathcal{E}\otimes P_i, \quad \Delta \mathcal{E}=\mathcal{E}\otimes\mathcal{E}\\
\Delta M_i=M_i\otimes 1+1\otimes M_i, & \quad \Delta N_i=N_i\otimes 1+\mathcal{E}\otimes N_i -\frac{1}{\kappa}\epsilon_{ijk}P_j\otimes M_k
\end{split}\end{equation}
and 
\begin{equation}\begin{split}
\varepsilon(P_0)&=\varepsilon(P_i)=\varepsilon(M_i)=\varepsilon(N_i)=0, \quad \varepsilon(\mathcal{E})=1\\
S(P_0)&=-P_0, \quad S(\mathcal{E})=\mathcal{E}^{-1}, \quad S(P_i)=-P_i \mathcal{E}^{-1}, \quad S(M_i)=-M_i\\
&S(N_i)=-\mathcal{E}^{-1}\left(N_i-\frac{1}{\kappa}\epsilon_{ijk}P_j M_k\right)
\end{split}\end{equation}
The algebra $\mathcal{M}_{\kappa}$ is a left-module over the Hopf algebra $\mathcal{P}_{\kappa}$ which can be viewed as the algebra of quantum symmetries of the corresponding noncommutative space. $\kappa$-Minkowski space\footnote{$\kappa$-Minkowski space viewed as the universal enveloping algebra of Lie algebra $\mathfrak{g}$ has a natural Hopf algebra structure with a primitive coproduct, antipode and counit.} is a dual of the Hopf subalgebra generated by $P_i$ and $\mathcal{E}$. The structure of $\mathcal{M}_{\kappa}$ as a left module over the Hopf algebra $\mathcal{P}_{\kappa}$ can be expressed for any $f\in\mathcal{M}_\kappa$ as 
\begin{equation}\begin{split}\label{djelovanje}
(\mathcal{E}\triangleright f)(x)=&f(x_0+\frac{i}{\kappa}, x_i), \quad (P_{\mu}\triangleright f)(x)=-i(\partial_{\mu}f)(x), \quad (M_i\triangleright f)=-i(\epsilon_{ijk}x_j \partial_k f)(x)\\
&(N_i\triangleright f)(x)=\left(\left[\frac{x_i}{2}(\kappa(1-\mathcal{E}^2)-\frac{1}{\kappa}\partial^{2}_{k})+i(x_0+\frac{1}{\kappa}x_k \partial_k)\partial_i\right]f\right)(x)
\end{split}\end{equation}
Notice that the generators of spatial translations are not derivations of the algebra $\mathcal{M}_{\kappa}$ due to the coproduct \eqref{cop}.\\

To define a well behaved field theory we need to find  some reasonable action functional.  It is natural to  impose the following three conditions on the physical action functional $\mathcal{S}_{\kappa}[\phi]$:
\begin{itemize}
\item The $\mathcal{P}_\kappa$-invariance of $\mathcal{S}_{\kappa}[\phi]$ \cite{sitarz, walletkappa, Agostini:2003vg}
\begin{equation}
h\triangleright \mathcal{S}_{\kappa}[\phi]=\varepsilon(h)\mathcal{S}_{\kappa}[\phi], \quad \forall h\in \mathcal{P}_\kappa
\end{equation}
\item The action $\mathcal{S}_{\kappa}[\phi]$ is positive and real.
\item $\mathcal{S}_{\kappa}[\phi]$ reduces to the standard scalar field theory in the commutative limit $\kappa\rightarrow\infty$.
\end{itemize}
In order to find an action $\mathcal{S}_{\kappa}[\phi]$ invariant under $\mathcal{P}_\kappa$ we consider functional  of the following form
\begin{equation}\label{form}
\mathcal{S}_{\kappa}[\phi]=\int d^4 x \mathcal{L}[\phi]
\end{equation}
where $\phi, \mathcal{L}[\phi]\in\mathcal{M}_{\kappa}$. Now, by using \eqref{djelovanje} one can show \cite{walletkappa}
\begin{equation}
P_{\mu}\triangleright \mathcal{S}_{\kappa}[\phi]=\int d^4 x P_{\mu}\triangleright \mathcal{L}[\phi]=0, \quad N_i\triangleright \mathcal{S}_{\kappa}[\phi]=0, \quad M_i\triangleright\mathcal{S}_{\kappa}[\phi]=0, \quad \mathcal{E}\triangleright\mathcal{S}_{\kappa}[\phi]=\mathcal{S}_{\kappa}[\phi]
\end{equation}
This way we have proven that the action of the form \eqref{form} fulfills the first condition and is invariant under the action of $\mathcal{P}_\kappa$.\\

Since the following properties hold \cite{walpol, Arzano:2014jfa}
\begin{equation}\label{pi}
\int d^4 x (f\star g^\dagger)(x)=\int d^4 x \ f(x)\bar{g}(x), \quad \int d^4 x\ f^\dagger (x)=\int d^4 x \ \bar{f}(x)
\end{equation}
it is easy to see
\begin{equation}\label{p}
\int d^4 x (f\star f^\dagger)(x)\geq 0, \quad \int d^4 x (f^\dagger \star f)(x)\geq 0
\end{equation}
which defines a positive map $\int d^4 x : \mathcal{M}_{\kappa^+}\rightarrow\mathbb{R}_+$ where $\mathcal{M}_{\kappa^+}$ denotes the set of positive elements of $\mathcal{M}_\kappa$. It is important to note that the Lebesgue integral does not define a cyclic trace. Namely
\begin{equation}
\int d^4 x (f\star g)(x)=\int d^4 x((\sigma\triangleright g)\star f)(x), \quad \sigma=e^{-\frac{3P_0}{\kappa}}.
\end{equation}
The operator $\sigma$ is an algebra automorphism often called twist. It gives rise to a twisted trace $\text{Tr}(a\star b)=\text{Tr}((\sigma\triangleright b)\star a)$ which signals the occurrence of a KMS condition. The operator $\sigma_t =e^{i\frac{3tP_0}{\kappa}}$ defines a group of automorphisms called the modular group of the KMS weight (for more details see \cite{walletkappa}).\\

In order to construct real action functional we use a natural Hilbert product\footnote{See Appendix for more properties} given by 
\begin{equation}\label{hp}
\left\langle f, g\right\rangle=\int d^4 x (f^\dagger\star g)(x)=\int d^4 x \bar{f}(x)(\sigma\triangleright g)(x), \quad \forall f,g\in\mathcal{M}_{\kappa}
\end{equation}
Notice that the Hilbert product \eqref{hp} is $\mathbb{R}$-valued for all $f,g\in\mathcal{M}_\kappa$ that satisfy $\left\langle f, g\right\rangle=\left\langle g, f\right\rangle$. Therefore, the reality condition for the action functional will be automatically  fulfilled if we use 
\begin{equation}\label{SK}
\mathcal{S}_{\kappa}[\phi]=\left\langle \phi, K\phi\right\rangle
\end{equation}
where $K$ is some self-adjoint kinetic operator satisfying $\left\langle \phi, K\phi\right\rangle=\left\langle K\phi,\phi\right\rangle$. It is straightforward to see that $P_{\mu}$ and $\mathcal{E}$ are self-adjoint with respect to the Hilbert product \eqref{hp}. In constructing the kinetic operator we assume it to be a pseudo-differential operator given by
\begin{equation}\label{K}
(K f)(x)=\int d^4 y d^4 p \ \tilde{K}(p)f(y)e^{ip(x-y)}
\end{equation}
for any $f$ in the domain of $K$ dense in the Hilbert space $\mathfrak{H}\cong L^{2}(\mathbb{R}^4)$, and $\tilde{K}(p)$ is the Fourier transform of the kernel of $K$. The self-adjointness of $K$ implies $\overline{\tilde{K}}(p)=\tilde{K}(p)$. \\

According to the discussion made so far and using \eqref{SK} and \eqref{K} we can finally  define the action functional that satisfies all three conditions by\footnote{One could also consider the mass term of the form $m^2\left\langle \phi, \phi\right\rangle$, but we will be focused just on the massless case when calculating the vacuum energy, because the mass term only affects the IR behavior of the theory.}
\begin{equation}\label{akcija}
\mathcal{S}_{\kappa}[\phi]=\left\langle \phi, K\phi\right\rangle=\int d^4 x d^4 y d^4 p \bar{\phi}(p)\phi(p)e^{ip(x-y)}e^{-\frac{3p_0}{\kappa}}\tilde{K}(p)
\end{equation}
If we compare this expression with \eqref{SF} we get that the relevant operator for calculating the vacuum energy is given by 
\begin{equation}\label{F}
F(x,y)=\int d^4 p e^{ip(x-y)}e^{-\frac{3p_0}{\kappa}}\tilde{K}(p) \quad \Longrightarrow \quad \tilde{F}(p)=e^{-\frac{3p_0}{\kappa}}\tilde{K}(p).
\end{equation}
So, in order to calculate vacuum energy in various $\kappa$-Poincar\'e invariant theories we  just need to extract the function $\tilde{F}(p)$. \\

We have a few natural choices for the explicit form of the kinetic operator in \eqref{akcija}. In this paper we will investigate three such choices: the Casimir operator, the square of modular Dirac operator and the equivariant Dirac operator.

The reason to choose the Casimir operator in the so called Majid-Ruegg \cite{majid} or bicrossproduct \cite{Lukierski:1992dt} basis is because it is a straightforward generalization of the usual Casimir operator of the Poincar\'e algebra and its corresponding momentum dispersion relations is one of the most studied in the literature.\\
The second example, i.e. modular operator, is motivated by the NC geometry \`a la Connes \cite{connes}.  In this approach the central object is the so called spectral triple where all the information about the space can be encoded in the triple $(A,\ \mathfrak{H},\  \mathcal{D})$, where $A$ is the algebra, $\mathfrak{H}$ the Hilbert space on which the algebra is represented and $\mathcal{D}$ the Dirac operator \cite{connes}. In \cite{modular} it was shown that if one wants to build a spectral triple for $\kappa$-Minkowski space one needs to relax some of the axioms for the corresponding Dirac operator, due to some boundlessness issue of the commutator between the Dirac operator and the elements of the algebra. In order to resolve the boundlessness issue one introduces a twisted commutator and a weight (related to the KMS structure \cite{walletkappa}) which replaces the usual trace in order to measure the growth of the resolvent of the Dirac operator. Then under some reasonable assumptions one can show that this operator is related to a unique Dirac operator with bounded twisted commutator, appropriate classical limit and a spectral dimension equal to the classical one \cite{modular}. \\
The third example, that is the equivariant Dirac operator  is interesting because it is singled out by the bicovariant differential calculus and also it has a typical pattern of dimensional reduction in which the spectral dimension decreases from the Hausdorff dimension (in the IR limit) to a smaller value in the UV regime \cite{Arzano:2014jfa}. This feature of dimensional reduction in the UV regime  is a common thread for a huge class of very different approaches to Quantum Gravity \cite{Carlip:2017eud}. This operator is also interesting because it is equivariant under the action of the quantum Euclidean group and emerges in the construction of equivariant spectral triple \cite{dandrea}. 

\section{Vacuum energy in $\mathcal{P}_{\kappa}$-invariant field theories}

We have set up everything for calculating the vacuum energy. All we need to do now is to evaluate \eqref{H} and \eqref{W} for different kinetic operators, that is \eqref{F}.

\subsection{The Casimir operator}
The Casimir operator $\mathcal{C}_{\kappa}$ commutes with all the generators of the $\kappa$-Poincar\'e algebra and as such we can use it as the kinetic operator. This choice fulfills all the conditions discussed in the previous section. Also this is in the complete analogy with the commutative case, where the Casimir operator of the Poincar\'e algebra serves as the kinetic operator. The Casimir operator $\mathcal{C}_{\kappa}$ in the Majid-Ruegg basis \cite{majid} is given by
\begin{equation}\label{C}
\mathcal{C}_{\kappa}(p)=4\kappa^2 \sinh^2\left(\frac{p_0}{2\kappa}\right)+e^{\frac{p_0}{\kappa}}p^{2}_{i}
\end{equation}
so that the function \eqref{F} that we need in order to calculate the vacuum energy is given by
\begin{equation}
\tilde{F}_{\mathcal{C}_{\kappa}}=e^{-\frac{3p_0}{\kappa}}\mathcal{C}_{\kappa}(p).
\end{equation}
The relevant integral in calculating the heat kernel \eqref{H} is given by
\begin{equation}\begin{split}\label{IC}
I_{\mathcal{C}_{\kappa}}(s)&=\int^{+\infty}_{-\infty}d^4 p \ e^{-s\tilde{F}_{\mathcal{C}_{\kappa}}}=4\pi\int^{+\infty}_{-\infty}dp_0 \ \exp\left[-s\kappa^2 e^{-\frac{3p_0}{\kappa}}\left(e^{\frac{p_0}{\kappa}}+e^{-\frac{p_0}{\kappa}}-2\right)\right]\int^{\infty}_{0}p^2dp\  e^{-sp^2e^{-\frac{2p_{0}}{\kappa}}} \\
&=\left(\frac{\pi}{s}\right)^{\frac{3}{2}}\int^{+\infty}_{-\infty}dp_0 \ \exp\left[\frac{3p_0}{\kappa}-s\kappa^2 e^{-\frac{3p_0}{\kappa}}\left(e^{\frac{p_0}{\kappa}}+e^{-\frac{p_0}{\kappa}}-2\right)\right]\\
&=\kappa\left(\frac{\pi}{s}\right)^{\frac{3}{2}}\int^{\infty}_{0}\frac{dy}{y^4}e^{-s\kappa^2\left(y^2+y^4-2y^3\right)}
\end{split}\end{equation}
where we used the integral $\int^{\infty}_{0} dx \ x^2 e^{-ax^2}=\frac{\sqrt{\pi}}{4a^{3/2}}$ in the first line and the substitution $y=e^{-\frac{p_0}{\kappa}}$ in the second line. The integral in \eqref{IC} is UV divergent, that is, it is singular for $y\rightarrow 0$. To illustrate this let us put a cutoff on the variable $p_0$ such that $\pm\infty\rightarrow\pm\Lambda$, then the integral we need to solve is given by  
\begin{equation}\label{limint}
\lim_{\Lambda\rightarrow\infty}\left(\int^{e^{\frac{\Lambda}{\kappa}}}_{e^{-\frac{\Lambda}{\kappa}}}\frac{dy}{y^4}e^{-s\kappa^2\left(y^2+y^4-2y^3\right)}\right)\approx\int^{e^{\frac{\Lambda}{\kappa}}}_{e^{-\frac{\Lambda}{\kappa}}}\frac{dy}{y^4}e^{-s\kappa^2 y^2}\longrightarrow\frac{1}{3}e^{\frac{3\Lambda}{\kappa}}+\mathcal{O}\left(e^{\frac{\Lambda}{\kappa}}\right)
\end{equation}
Now, one has to be careful with the regularization and interpreting the cutoff in $p_0$, that is $\Lambda$ as the UV limit. Namely, one has to take in account the $\kappa$-Poincar\'e symmetry. Motivated by the Hopf algebra structure of the $\mathcal{P}_\kappa$, especially the deformed translation algebra generated by $P_i$ and $\mathcal{E}$ (see \eqref{cop}), it is more natural to interpret $y=e^{-\frac{p_0}{\kappa}}$ as related to the physical quantity replacing $p_0$ in the NCFT. More precisely, the Casimir operator \eqref{C} can be written as
\begin{equation}
\mathcal{C}_{\kappa}=e^{\frac{p_0}{\kappa}}\left(\mathcal{P}^{2}_{0}+p^{2}_{i}\right), \quad \mathcal{P}_{0}=\kappa(1-y)
\end{equation}
so that the relevant quantity for the $\mathcal{P}_\kappa$-covariant quantum field theories, which reduces to $p_0$ in the commutative limit is given by $\mathcal{P}_{0}$. Putting a cutoff $\left|\mathcal{P}_{0}\right|\leq M$ enables us to derive the appropriate cutoff on the variable $p_0$ 
\begin{equation}\label{cutoff}
\Lambda=\kappa\ln\left(1+\frac{M}{\kappa}\right).
\end{equation}
Taking the relation between the cutoffs \eqref{cutoff} and  \eqref{limint}, the integral in \eqref{IC} and the corresponding effective action $W_{\mathcal{C}_{\kappa}}$ reads
\begin{equation}
I_{\mathcal{C}_{\kappa}}(s)=\frac{\kappa}{3}\left(\frac{\pi}{s}\right)^{\frac{3}{2}}\left(1+\frac{M}{\kappa}\right)^3, \quad \Longrightarrow \quad W_{\mathcal{C}_{\kappa}}=-\frac{\kappa V}{144\pi^{5/2}}M^3 \left(1+\frac{M}{\kappa}\right)^3
\end{equation}
 so that the vacuum energy becomes
\begin{equation}
\rho_{vac}(M)=\frac{\kappa}{144\pi^{5/2}}M^3 \left(1+\frac{M}{\kappa}\right)^3
\end{equation}
We see that the vacuum energy behaves as $\rho_{vac}\propto\frac{M^6}{\kappa^2}$ in the UV limit, which is more divergent when compared with the commutative case. In case that the deformation parameter is equal to the UV cutoff $\kappa=M$ we recover the commutative behavior\footnote{Namely, one does not need to set the cut-off to the deformation parameter scale $\kappa$ in order to recover the commutative limit.  In doing so, that is by putting $M=\kappa$ one is left with only one dimensionfull parameter in the theory, $\kappa$, and by dimensional analysis the leading term has to be $\rho\propto(\text{mass scale})^4$. So in the commutative case this scale is provided by the cut-off, and if cut-off is set to be $\kappa$ then we get the same behavior which is then just confirmed by our  explicit calculation followed by setting the limit $M=\kappa$ in the end.
The commutative limit is given by $\kappa\longrightarrow\infty$, one just has to be careful  about one subtle point, and this is the fact that the limits $\kappa\longrightarrow\infty$ and $M\longrightarrow\infty$ may not commute due to divergences and in general even possible UV/IR mixing, so the safe way of doing the commutative limit is setting $\kappa\longrightarrow\infty$ at the initial stage (like in eq.\eqref{F}) calculate the lowest order of the NC correction and then perform the calculation of the integrals and set the $M\longrightarrow\infty$ limit (like in \cite{andjelo, andmim}).} $\rho_{vac}\propto M^{4}$. The result obtained here is in agreement with the investigation performed in \cite{andjelo, andmim} where, even though there they used a different integration measure,  the same behavior for the vacuum energy was discovered.\\

\subsection{Modular operator}
 Another choice for the kinetic operator comes from studying modular spectral triples \cite{modular}. It is related to the Casimir operator by
\begin{equation}
\mathcal{M}(p)=e^{-\frac{p_0}{\kappa}}\mathcal{C}_{\kappa}(p)=\kappa^2\left(1-e^{-\frac{p_0}{\kappa}}\right)^2+p^{2}_{i}
\end{equation}
so that the relevant function is given by $\tilde{F}_{\mathcal{M}}=e^{-\frac{3p_0}{\kappa}}\mathcal{M}(p)$ and the integral of interest is
\begin{equation}
I_{\mathcal{M}}(s)=\int^{+\infty}_{-\infty}d^4 p \ e^{-s\tilde{F}_{\mathcal{M}}}=4\kappa\left(\frac{\pi}{s}\right)^{\frac{3}{2}}\int^{\infty}_{0}\frac{dy}{y^{\frac{9}{2}}}e^{-s\kappa^2(y^3-2y^4+y^5)}\approx 4\kappa\left(\frac{\pi}{s}\right)^{\frac{3}{2}}\int^{\infty}_{0}\frac{dy}{y^{\frac{9}{2}}}e^{-s\kappa^2 y^3}
\end{equation}
where we again use the same strategy as in the previous subsection. The integral in question is UV divergent so we regularize it and keep only the most divergent part, keeping in mind \eqref{cutoff}
\begin{equation}\label{ikk}
\lim_{\Lambda\rightarrow\infty}I(s)=4\kappa\left(\frac{\pi}{s}\right)^{\frac{3}{2}}\int^{e^{\frac{\Lambda}{\kappa}}}_{e^{-\frac{\Lambda}{\kappa}}}\frac{dy}{y^{\frac{9}{2}}}e^{-s\kappa^2y^3}\approx 4\kappa \left(\frac{\pi}{s}\right)^{\frac{3}{2}}\frac{(s\kappa^2)^{7/6}}{3}\Gamma\left(-\frac{7}{6}, s\kappa^2 e^{-\frac{3\Lambda}{\kappa}}\right)\approx \frac{8\kappa}{7}\left(\frac{\pi}{s}\right)^{\frac{3}{2}}\left(1+\frac{M}{\kappa}\right)^{\frac{7}{2}}
\end{equation}
where we used the expansion for the incomplete Gamma function $\Gamma\left(-\frac{7}{6}, ax^3\right)\approx \frac{6}{7(ax)^{7/6}}+\mathcal{O}(x^{-1/2})$. Now, after taking into account \eqref{W}, \eqref{H} and \eqref{ikk} we get for the vacuum energy
\begin{equation}
\rho_{vac}=\frac{\kappa}{42\pi^{\frac{5}{2}}}M^{3}\left(1+\frac{M}{\kappa}\right)^{\frac{7}{2}}
\end{equation} The corresponding vacuum energy is $\rho_{vac}\propto\sqrt{\frac{M^{13}}{\kappa^5}}$ which is even more divergent that in the case of Casimir operator. In the limit $\kappa=M$ we recover the commutative behavior.\\

\subsection{Equivariant Dirac operator}

In this subsection we investigate the square of the equivariant Dirac operator. This operator comes from the construction of the equivariant spectral triple for the $\kappa$-Minkowski space \cite{dandrea} and is related to the Casimir operator by
\begin{equation}
\mathcal{K}_{\kappa}(p)=\mathcal{C}_{\kappa}(p)+\frac{1}{4\kappa^2}\mathcal{C}^{2}_{\kappa}(p)
\end{equation}
The relevant function is given by $\tilde{F}_{\mathcal{K}}(p)=e^{-\frac{p_0}{\kappa}}\mathcal{K}_\kappa$ and for the sake of future convince we write it in the following form
\begin{equation}
\tilde{F}_{\mathcal{K}}(p)=A(p_0)+B(p_0)p^{2}_{i}+C(p_0)p^{4}_{i}
\end{equation}
where
\begin{equation}
A=\frac{\kappa^2}{4}(y+2y^2-2y^3+2y^4+y^5), \quad B=\frac{1}{2}(y+y^3), \quad C=\frac{y^2}{4\kappa^2}, \quad y=e^{-\frac{p_0}{\kappa}}
\end{equation}
The integral we need to obtain in order to calculate the heat kernel is given by
\begin{equation}\label{IK}
I_{\mathcal{K}}(s)=\int^{+\infty}_{-\infty}d^4 p \ e^{-s\tilde{F}_{\mathcal{K}}}=4\pi\int^{+\infty}_{-\infty}dp_0 \ e^{-sA}\int^{\infty}_{0}p^2dp\  e^{-s(Bp^2+Cp^4)} 
\end{equation}
We will first deal with the integral over the radial variable $p$ and for that we need the following integral
\begin{equation}
\int^{\infty}_{0}x^2dx\  e^{-(ax^2+bx^4)}=\frac{1}{4b^{5/4}}\left[\sqrt{b}M\left(\frac{3}{4},\frac{1}{2},\frac{a^2}{4b}\right)-a\Gamma\left(\frac{5}{4}\right)M\left(\frac{5}{4},\frac{3}{2},\frac{a^2}{4b}\right)\right] 
\end{equation}
where $M(a,b,z)$ is the confluent hypergeometric function. Since we will be interested in the the most divergent parts we look at  the limit $s\longrightarrow 0$  where we have\footnote{Where we used that the confluent hypergeometric function satisfies $M(a,b,0)=1$} $\int^{\infty}_{0}p^2dp\  e^{-s(Bp^2+Cp^4)}\approx \frac{1}{4(sC)^{3/4}}+\mathcal{O}(s^{-1/4}) $ so that \eqref{IK} becomes
\begin{equation}
I_{\mathcal{K}}(s)\approx\pi\left(\frac{4\kappa^2}{s}\right)^{\frac{3}{4}}\int^{+\infty}_{-\infty}dp_0 \ e^{\frac{3p_0}{\kappa}-sA}=\kappa\pi\left(\frac{4\kappa^2}{s}\right)^{\frac{3}{4}}\int^{\infty}_{0}\frac{dy}{y^{\frac{5}{2}}}e^{-sA}
\end{equation}
The integral in \eqref{IK} is once again divergent and therefore we will proceed with the same strategy as in the former cases and use the cutoff regularization procedure. In doing so, we get
\begin{equation}\label{ik}
\lim_{\Lambda\rightarrow\infty}I_{\mathcal{K}}(s)=\kappa\pi\left(\frac{4\kappa^2}{s}\right)^{\frac{3}{4}}\int^{e^{\frac{\Lambda}{\kappa}}}_{e^{-\frac{\Lambda}{\kappa}}}\frac{dy}{y^{\frac{5}{2}}}e^{-sA}\approx\kappa\pi\left(\frac{4\kappa^2}{s}\right)^{\frac{3}{4}}\int^{e^{\frac{\Lambda}{\kappa}}}_{e^{-\frac{\Lambda}{\kappa}}}\frac{dy}{y^{\frac{5}{2}}}e^{-\frac{s\kappa^2}{4}y}=\frac{2\kappa\pi}{3}\left(\frac{4\kappa^2}{s}\right)^{\frac{3}{4}}e^{\frac{3\Lambda}{2\kappa}}+\text{finite part}
\end{equation}
Now, after taking into account \eqref{W}, \eqref{H} and \eqref{ik} we get for the vacuum energy
\begin{equation}\label{lambdaD}
\rho_{vac}=\frac{4}{9\pi^3}\left(\frac{\kappa}{2}\right)^{\frac{5}{2}}M^{\frac{3}{2}}\left(1+\frac{M}{\kappa}\right)^{\frac{3}{2}}
\end{equation}
The vacuum energy for the square of the equivariant Dirac operator has a quite better behavior then the commutative theory since $\rho_{vac}\propto \kappa M^{3}$. This improvement in the UV behavior of the  quantum field theory was already demonstrated in \cite{walletkappa} where the properties of the 2-point function was investigated, and also in \cite{walpol} where it was shown that the $\beta$-function for the various versions of NC $\phi^4$-theory vanishes. In the limit $\kappa=M$ we recover the usual commutative behavior.\\

%%%%%%%%%%%%%%%%%%%%%%%%%%%%%%%%%%%%%%%%%%%%%%%%%%%%%%%%%%%%%%%%%%%%%%%%%
%%%%%%%%%%%%%%%%%%%% %%%%%%%%%%%%%%%%%%%%%
%%%%%%%%%%%%%%%%%%%%%%%%%%%%%%%%%%%%%%%%%%%%%%%%%%%%%%%%%%%%%%%

\section{Final remarks}
The quantum fluctuations of the vacuum contribute to the expectation value of the energy-momentum tensor in a way that mimics the cosmological constant. In QFT the vacuum energy is highly nontrivial, as in the case of a simple quantum harmonic oscillator in the ground state, each mode of every field has its own contribution to the zero-point energy. This energy arises from virtual particle-antiparticle pairs, i.e. loops. The corresponding energy-momentum tensor can be written as $\left\langle T_{\mu\nu}\right\rangle=-\rho_{vac}g_{\mu\nu}$. Even though it would appear on the right hand side of the Einstein's equation, vacuum energy has the form of a cosmological constant, therefore we can absorb it in order to redefine $\lambda\longrightarrow\lambda_{ren}=\lambda +8\pi G \rho_{vac}$. Equivalently, we may absorb the bare cosmological constant  $\lambda$ appearing in the Einstein's equation
into the energy density of the vacuum \cite{Bousso:2007gp}. This way one turns the cosmological constant problem into a fine tunning one. 

The novelty that the noncommutativity brought to this problem is the fact that the UV behavior of the theory is changed, resulting in a different behavior, that is  change in the level of divergence of the vacuum energy. This, of course,  is just a starting point and one should proceed with renormalization of the vacuum energy in order to absorb the cut-off dependance into the counterterms \cite{Grensing}. This is an adventure of its own, since the quantum corrections are non trivial \cite{walletkappa, walpol} and furthermore the counterterm should be steaming from the cosmological constant part of the gravitational action. Now, in the NC setting even this procedure should be in some sense modified since the gravitational part also acquires NC correction which would give more possible counterterms that are similar to the cosmological term \cite{Harikumar:2006xf}-\cite{Borowiec:2016zrc}. This we plan to investigate in future works.
\\

In this work we have presented a construction of $\kappa$-Poincar\'e invariant field theory and investigated the vacuum energy for several choices of kinetic operator. 

We shown that the behavior of the vacuum energy highly depends on the choice of the kinetic operator. Our  guiding principles, in how to  ``naturally'' select a kinetic operator are mostly governed by arguments from NC geometry. Namely,  as illustrated in Section III, we extensively used the properties of the $\kappa$-Minkowski algebra in our construction of the field theory. First we constructed the $\star$-product by adopting the Wigner-Weyl quantization scheme in III.a, and then in III.b we  exploited the quantum symmetries of $\kappa$-Minkowski space, i.e. $\kappa$-Poincare-Hopf algebra, in order to construct allowable candidates for a physical action.  We also imposed various other conditions motivated from physics and mathematics (reality, positivity, commutative limit, etc.)  Unfortunately all  this procedures does not single out a unique  action, so we were forced to choose some examples basically as an ``educated guess'' steaming from NC geometry. Even though our three examples (Casimir, modular and equivariant Dirac operator) are well motivated from NC geometry point of view, the question of the  ``right'' choice of  the kinetic operator, even after imposing $\kappa$-Poincare symmetry, it is still an open one because it depends heavily on the choice of the differential calculus or spectral triple one uses.

 It is shown that in the case of the square of equivariant Dirac operator one obtains an improved UV behavior of the zero-point energy. Therefore one needs to revise the cosmological constant problem. 
Even though in the case of the square of the equivariant Dirac  we obtained a significant improvement in the UV behavior it is still not sufficient to resolve the cosmological constant problem. Namely, if we would assume that the UV cutoff is proportional to the Planck mass $M=M_{Pl}\approx 10^{18}$GeV, and compared the theoretical estimate obtained in \eqref{lambdaD} with the observed one $\rho^{obs}_{vac}\approx 10^{-47}$GeV$^4$, then we could get a bound on the deformation parameter $\kappa$ that would be far too low, indicating that we should have already observed NC effects in various experiments. However, one can speculate that the new energy scale, governed by the deformation parameter $\kappa$, is in the interval $\kappa\in[1$TeV$, \ M_{Pl}]$. This assumption is supported by the fact that more or less above the energy of 1TeV we did not explore particle interactions experimentally. In doing so, we would get a discrepancy with the observations ranging between 10 and 120 orders of magnitude. Therefore one is forced to understand the expression \eqref{lambdaD} as a starting point of the renormalization procedure, i.e. as part that will be reabsorbed by the counter-term. In doing so one transforms the cosmological constant problem into a fine tuning one, but with a sufficiently better starting point for tuning since it was shown that the quantum corrections in the NC version of the $\phi^4$ theory are finite, namely the $\beta$-function vanishes perturbatively \cite{walpol}.\\

When introducing curvature in consideration then things get much more complicated and interesting \cite{Guberina:2007fa}.  One encounters the so called running cosmological constant \cite{Shapiro:2003kv}. In order to provide a small value of the observed cosmological constant one can also introduce the vacuum term which cancels the induced one at some point in the very far infrared cosmic scale \cite{Shapiro:1999zt}. Namely, one can even show that the renormalized vacuum energy in curved background is $\rho_{vac}\approx m^2 H^2$ , where $m$ is the mass of the scalar field and $H$ the Hubble scale \cite{Kohri:2016lsj}. Therefore one is tempted to extend the aforementioned ideas  to the NC setting. We leave this line of research for future work.\\

There have been other investigations concerning NC effects on the cosmological constant problem. The implications to cosmology steaming from the spectral triple approach to NC geometry was reported in \cite{Marcolli}. In \cite{Garattini:2010dn, Garattini:2012km} the study of the cosmological constant problem as an eigenvalue problem of a certain Sturm-Liouville problem was performed. The authors of \cite{Garattini:2010dn, Garattini:2012km} employed the NC effects into the Weeler-De Witt equation through minimal length. It would be very interesting and challenging to accommodate the two aforementioned approaches for  the case of  $\kappa$-Poincar\'e symmetry and $\star$-product for $\kappa$-Minkowski space. \\ 

One has to be aware that the calculations performed in this paper are within the framework of NC field theory, so we are really only able to comment on the NC effect on the vacuum or zero-point energy and not on  the whole cosmological constant problem. Namely, noncommutativity also affects the geometric or ``pure'' gravity side of the Einstein's equation. In numerous approaches \cite{Harikumar:2006xf, Aschieri:2005yw, Ciric:2016ycq, Aschieri:2007eq, Chamseddine:1992yx, Calmet:2005qm, Franco:2009hc, Ohl:2009pv} we have witnessed the NC corrections to Einstein's equation. Moreover, in \cite{Beggs:2013pxa, Majid:2014afa, Borowiec:2016zrc} it has been indicated that the nature of the cosmological constant could be entirely noncommutative. Therefore one needs to find the proper NC correction to the geometric side of the Einstein's  equation and then together with the result obtained in this paper investigate further all the consequences.\\

One of the most obvious manifestation of vacuum energy (besides the cosmological constant) is the so called Casimir effect \cite{casimir}. It would be very interesting to calculate the modification of the scalar Casimir force between two parallel plates due to noncommutativity of the spacetime and  compare it with the existing results in the literature \cite{Khelili:2011pv, Casadio:2007ec, Demetrian:2002pm, Chaichian:2001pw, Nouicer:2009ze}.\\

Finally, we can conclude that the work done in this paper does not solve the cosmological constant problem, but rather it shows that the UV behavior of the vacuum energy, and therefor cosmological constant, can be improved. This is a first step (at least as far as $\kappa$-deformations are concerned) toward approaching the cosmological constant problem  that gives some properties of the nature of the NC field theories.  It still remains to perform renormalization, but that requires a careful inspection of the possible conterterms coming from the NC corrections to the Einstein-Hilbert action, because this might lead  to some convenient cancellations. Also, one has to note that this is still very model-dependent, that is even though we have enforced the $\kappa$-Poincare symmetry from the start we still have a plethora of possible candidates for the kinetic operator to choose from. We have presented only three of them, from which the equivariant Dirac operator showed an improvement in the UV behavior, but it is still not excluded that there is some other choice that would steam from some other differential calculus or (twisted) spectral triple that could indeed render a finite value for the vacuum energy.\\

\noindent{\bf Acknowledgment}\\
T.J. is very grateful to  Andjelo Samsarov for  numerous and fruitful discussions and to Ivica Smoli\'c for comments. 
This work was partially supported by the H2020 CSA Twinning project No. 692194, RBI-T-WINNING and  by the European Union through the European Regional Development Fund - the Competitiveness and Cohesion Operational Programme (KK.01.1.1.06).\\

\appendix
\section{$\kappa$-Minkowski space and the $\star$-product}
$\kappa$-Minkowski space can be viewed as the universal enveloping algebra of the Lie algebra $\mathfrak{g}$ defined by
\begin{equation}\label{g}
[x_0,x_i]=\frac{i}{\kappa}x_i, \quad [x_i,x_j]=0, \quad i,j=1,...,d
\end{equation}
where $\kappa$ is the deformation parameter usually related to Planck mass or some other quantum gravity scale. The coordinates $x_0$ and $x_i$ are self-adjoint operators acting on a suitable Hilbert space. Since the derived algebra $[\mathfrak{g},\mathfrak{g}]$ is nilpotent it follows that the Lie algebra $\mathfrak{g}$ is solvable and therefore the associated Lie group $\mathcal{G}$ is also solvable. In order to investigate further the $\kappa$-Minkowski space it is useful to use the framework of group algebras and $\mathcal{C}^{\star}$-dynamical systems \cite{sitarz, cstar}. Notice that the group $\mathcal{G}$ is not unimodular and therefore we have a distinct left and right invariant Haar measure, which we denote by $d\mu$ and $d\nu$ respectively. These measures are related by  a modular function $d\nu(s)=\Delta_{\mathcal{G}}(s)d\mu(s)$. For the sake of simplicity let us consider $d=1$ case. In doing so, we see that the Lie group $\mathcal{G}$ is completely characterized by defining
\begin{equation}\label{w}
W(p_0,p_1)=e^{ip_1 x_1}e^{ip_0 x_0}
\end{equation}
where $p_0$, $p_1 \in \mathbb{R}$ are coordinates on the group manifold and can be interpreted as momenta. Using \eqref{g} and \eqref{w} one obtains the group law for the multiplication in the group $\mathcal{G}$
\begin{equation}
W(p_0,p_1)W(q_0,q_1)=W(p_0+q_0, p_1+e^{-\frac{p_0}{\kappa}}q_1)
\end{equation}
It is easy to see that the unit element $\mathbb{I}_{\mathcal{G}}$ and the inverse $W^{-1}$ are given by
\begin{equation}
\mathbb{I}_{\mathcal{G}}=W(0,0), \quad W^{-1}(p_0,p_1)=W(-p_0, -e^{\frac{p_0}{\kappa}}p_1)
\end{equation}
 Notice that the $\star$-product of 2-dim $\kappa$-Minkowski algebra can be obtained using Weyl quantization \cite{weyl, phasespace}. To illustrate this we have to start  from $L^1 (\mathcal{G})$, that is the convolution algebra of $\mathcal{G}$ and a representation $\pi_u :\mathcal{G}\longrightarrow\mathcal{B}(\mathfrak{H})$ that is a (strongly continuous) unitary representation of $\mathcal{G}$, where $\mathfrak{H}$ is a suitable Hilbert space and $\mathcal{B}(\mathfrak{H})$ is the $\mathcal{C}^{*}$-algebra of bounded operators on $\mathfrak{H}$. We define the convolution algebra with respect to the right invariant measure as $\hat{\mathcal{G}}=(L^1 (\mathcal{G}), \circ)$ where the algebra multiplication is the convolution defined by
\begin{equation}\label{conv}
(f\circ g)(t)=\int_{\mathcal{G}}d\nu(s)f(ts^{-1})g(s), \quad \forall \ t\in\mathcal{G}, \ f, \ g\in L^1 (\mathcal{G}).
\end{equation}
The unitary representation of the convolution algebra $\hat{\mathcal{G}}$ is given by $\pi:L^1 (\mathcal{G})\longrightarrow\mathcal{B}(\mathfrak{H})$ and
\begin{equation}\label{pii}
\pi(f)=\int_{\mathcal{G}}d\nu(s)f(s)\pi_u (s)
\end{equation}
so that it is a non-degenerate $*$-representation. One can easily check that
\begin{equation}
\pi(f)^\dagger=\pi(f^*), \quad \pi(f\circ g)=\pi(f)\pi(g), \quad \forall f,g\in\hat{\mathcal{G}}
\end{equation}
where the $\dagger$ denotes the adjoint operation acting on the operators with respect to the Hilbert product \eqref{hp}
\begin{equation}
\left\langle u, \ \pi(f)^\dagger  v\right\rangle=\left\langle \pi(f)u,\ v\right\rangle=\int_{\mathcal{G}}d\nu(s)\overline{f}(s)\left\langle u, \ \pi_u (s^{-1}) v\right\rangle
\end{equation}

In order to obtain the $\star$-product we will use the Weyl quantization map \cite{weyl}. This way one makes the identification between the functions on $\mathcal{G}$ with functions on $\mathbb{R}^2$. First, let
\begin{equation}
\mathcal{F}[f(p_0,p_1)]=\int_{\mathbb{R}^2}dx_0 dx_1 \ e^{-i(x_0 p_0 +x_1 p_1)}f(x_0,x_1)
\end{equation}
be the Fourier transform of $f\in L^1 (\mathbb{R}^2)$, then the quantization map for any $f\in L^1 (\mathbb{R}^2)\cap\mathcal{F}^{-1}[L^1 (\mathbb{R}^2)] $ is defined by
\begin{equation}
Q(f)=\pi\left(\mathcal{F}[f]\right)
\end{equation}
where $\pi$ is the unitary representation given by \eqref{pii}. Since the quantization map $Q$ must be a morphism of algebra, one writes
\begin{equation}
Q(f\star g)=Q(f)Q(g)=\pi\left(\mathcal{F}[f]\right)\pi\left(\mathcal{F}[g]\right)=\pi\left(\mathcal{F}[f]\circ\mathcal{F}[g]\right)
\end{equation} 
therefore giving us the expression for the $\star$-product
\begin{equation}
f\star g=\mathcal{F}^{-1}\left[\mathcal{F}[f]\circ \mathcal{F}[g] \right]
\end{equation}
and similarly one finds
\begin{equation}
f^\dagger = \mathcal{F}^{-1}\left[\mathcal{F}[f]^*\right]
\end{equation}
Finally, by using the fact that the right invariant measure is $d\nu (p_0,p_1)=dp_0 dp_1$ and the expression for right convolution product \eqref{conv}, one obtains
\begin{equation}
(f\star g)(x_0, x_1)=\int dp_0 dy_0 \  e^{-iy_0 p_0}f(x_0+y_0, x_1)g(x_0, e^{-\frac{p_0}{\kappa}}x_1)
\end{equation}
and
\begin{equation}
f^{\dagger}(x_0,x_1)=\int dp_0 dy_0 \ e^{-iy_0 p_0}\overline{f}(x_0+y_0, e^{-\frac{p_0}{\kappa}}x_1 )
\end{equation}
for any $f$ and $g\in \mathcal{F}[\mathcal{S}_c]$, where $\mathcal{S}_c$ denotes the space of Schwartz functions on $\mathbb{R}^2$ with compact support. 
\\

\section{Properties of the Hilbert product}
We introduce the following Hilbert product
\begin{equation}\label{hpa}
\left\langle f, g\right\rangle=\int d^4 x (f^\dagger\star g)(x)=\int d^4 x \bar{f}(x)(\sigma\triangleright g)(x), \quad \forall f,g\in\mathcal{M}_{\kappa}
\end{equation}
The positivity of the Hilbert product \eqref{hpa} is a consequence of \eqref{p}, while $\overline{\left\langle f, g\right\rangle}=\left\langle g,  f\right\rangle$ is guaranteed by \eqref{pi}. The related Hilbert space $\mathfrak{H}$ can be obtained canonically from the GNS construction by completining the linear space $\mathcal{M}_{\kappa}$ with respect to the norm
\begin{equation}
\left\|f\right\|^2=\left\langle f,  f\right\rangle=\int d^4 x (f^\dagger\star f)(x)=\int d^4 x \left|f^\dagger (x)\right|^2
\end{equation}
There is a unitary equivalence between $\mathfrak{H}$ and $L^2(\mathbb{R}^4)$ given by the intertwining map $\mathcal{A}_{\kappa}: \mathcal{M}_{\kappa}\rightarrow L^2(\mathbb{R}^4)$ defined by 
\begin{equation}
(\mathcal{A}_{\kappa} f)(x)=\int dp_0dy_0 \ e^{iy_0 p_0} f(x_0 +y_0, e^{-\frac{p_0}{\kappa}}x_i), \quad \overline{(\mathcal{A}_{\kappa}f)}(x)=f^{\dagger}(x)
\end{equation}
and it immediately follows 
\begin{equation}
\left\|\mathcal{A}_{\kappa}f\right\|^{2}_{2}=\int d^4 x \overline{(\mathcal{A}_{\kappa} f)}(x)(\mathcal{A}_{\kappa} f)(x)=\int d^4 x \left|f^\dagger (x)\right|=\left\|f^\dagger\right\|^2
\end{equation}
Hence, $\mathcal{A}_{\kappa}$ defines an isometry which extends to $\mathfrak{H}\rightarrow L^{2}(\mathbb{R}^4)$ while the surjectivity of $\mathcal{A}_{\kappa}$ steams directly from the existence of $\mathcal{A}^{-1}_{\kappa}$ together with density of $\mathcal{M}_{\kappa}$ in $L^{2}(\mathbb{R}^4)$ (for more details see \cite{walletkappa}).\\

%%%%%%%%%%%%%

\end{document}